\documentclass[sigconf,table,prologue]{acmart}

\AtBeginDocument{%
	\providecommand\BibTeX{{%
			\normalfont B\kern-0.5em{\scshape i\kern-0.25em b}\kern-0.8em\TeX}}}

\setcopyright{acmcopyright}
\copyrightyear{2019}
\acmYear{2019}

\acmConference{Humans in the Loop: Enabling and Facilitating Research on Cloud Computing at PEARC 19}{July 29th, 2019}{Chicago, IL}
\acmBooktitle{Humans in the Loop: Enabling and Facilitating Research on Cloud Computing at PEARC 19, July 29th, 2019, Chicago, IL}
\acmPrice{15.00}
\acmISBN{978-1-4503-9999-9/18/06}

\usepackage{listings}

\begin{document}
	
	\title{Enabling Microsoft OneDrive Integration with HTCondor}
	
	\author{Derek Weitzel}
	\email{dweitzel@unl.edu}
	\orcid{0000-0002-8115-7573}
	\affiliation{%
		\institution{University of Nebraska - Lincoln}
		\city{Lincoln}
		\state{NE}
	}

	\renewcommand{\shortauthors}{Weitzel}
	
\begin{abstract}
Accessing data from distributed computing is essential in many workflows, but can be complicated for users of cyberinfrastructure.  They must perform multiple steps to make data available to distributed computing using unfamiliar tools.  Further, most research on data distribution has focused on the efficiency of providing data to computing resources rather than considering the ease of use for distributing data.  Creating an easy to use data distribution method can reduce the time researchers spend learning cyberinfrastructure and increase its usefulness.

Microsoft OneDrive is a online storage solution providing both file storage and sharing.  OneDrive provides many different clients to access data stored in the service.  It provides many features that users of cyberinfrastructure could find useful such as automatic synchronization with desktop clients.

A barrier to using services such as OneDrive is the credential management neccessary to access the service.  Recent innovations in HTCondor have allowed the management of OAuth credentials to be handled by the scheduler on the user's behalf.  The user no longer has to copy credentials along with the job, HTCondor will handle the acquisition, renewal, and secure transfer of credentials on the user's behalf.

In this paper, I will focus on providing an easy to use data distribution method utilizing Microsoft OneDrive.  Measuring ease of use is difficult, therefore I will will describe the features and advantages of using OneDrive.  Additionally, I will compare it to measurements of data distribution methods currently used on a national cyberinfastructure, the Open Science Grid.

\end{abstract}

	\maketitle
	
	\section{Introduction}

Data distribution on cyberinfrastructure, such as the Open Science Grid (OSG) \cite{pordes2007open}, is a requirement for most workflows. For example, on the OSG , many users utilize the StashCache \cite{weitzel2019stashcache} data distribution federation or HTTP proxies.  Large experiments such as CMS \cite{chatrchyan2008cms} or ATLAS \cite{aad2008atlas} have dedicated software, services, and personnel dedicated to the distribution of data for processing.  Smaller organization or individual users do not have the expertise or manpower to distributed data on a heterogeneous cyberinfrastructure.

\begin{figure}[ht]
\includegraphics[width=0.45\textwidth]{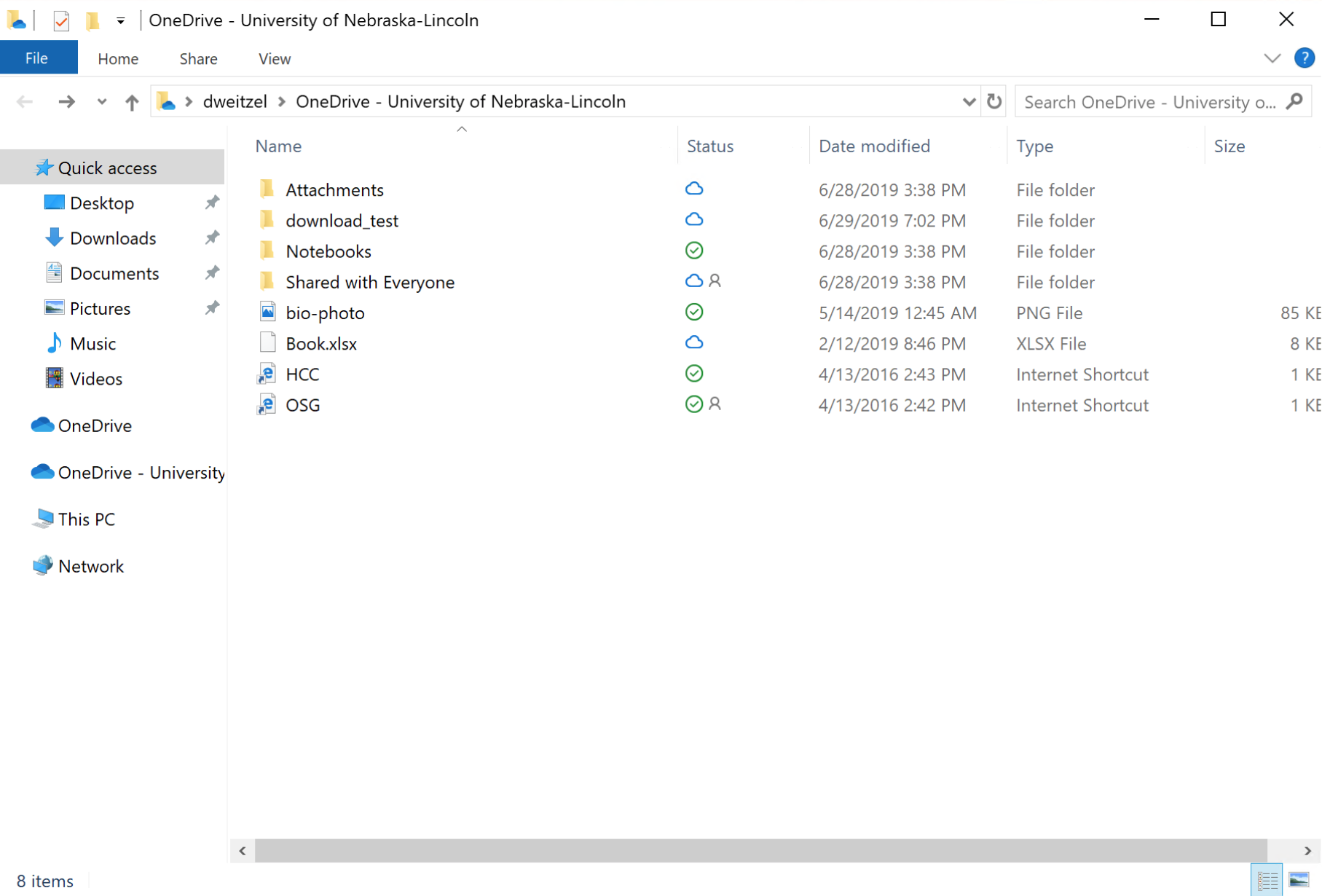}
\caption{Windows OneDrive Folder View}
\label{fig:onedrivefolder}
\end{figure}

Microsoft OneDrive \cite{onedrive} is a online storage solution providing both file storage and sharing.  OneDrive provides many different clients to access data stored in the service.  It provides many features that users of cyberinfrastructure could find useful such as automatic synchronization with desktop clients.  Users can simply copy input files on their desktop to the OneDrive folder and it is made available.  On Windows, the OneDrive folder is presented as a normal directory, as shown in Figure \ref{fig:onedrivefolder}.

Users are familiar with accessing and editing files from within their own desktop's folder view.  Other data distribution techniques require copying the file to separate storage services with unfamiliar tools such as \texttt{scp} or \texttt{rsync} \cite{tridgell1996rsync}.  Additionally, OneDrive is provided along with a Office 365 account, which many univerities have subscribed.

Until recently, users of OneDrive or similar services where required to manage their own credentials.  This required copying privileged keys with the jobs to multiple sites around the world.  This creates a security issue when these keys are transferred to untrusted sites.  

HTCondor recently added the capability to manage OAuth tokens on behalf of users \cite{withers2018scitokens}.  HTCondor will handle the credential acquisition and renewal on behalf of users.  Additionally, the credentials sent with jobs have short lifetimes, therefore providing a level of security on untrusted resources.

In this paper, I will discuss providing an easy to use data distribution method utilizing Microsoft OneDrive \cite{onedrive}.  Measuring ease of use is difficult, therefore I will will describe the features and advantages of using OneDrive.  I will also compare OneDrive to measurements of data distribution methods currently used on a national cyberinfrastructure, the Open Science Grid.






\section{Background}

HTCondor \cite{thain2005distributed} is a high throughput distributed computing manager.  It recently has added the capability to retrieve and manage OAuth credentials \cite{withers2018scitokens}.  A deamon in HTCondor, the CredMon, acts as an OAuth client which will guide the users through acquiring their credentials.  The CredMon sends the credentials to the HTCondor system for secure transfer to the execution host.

\begin{figure}[ht]
\includegraphics[width=0.4\textwidth]{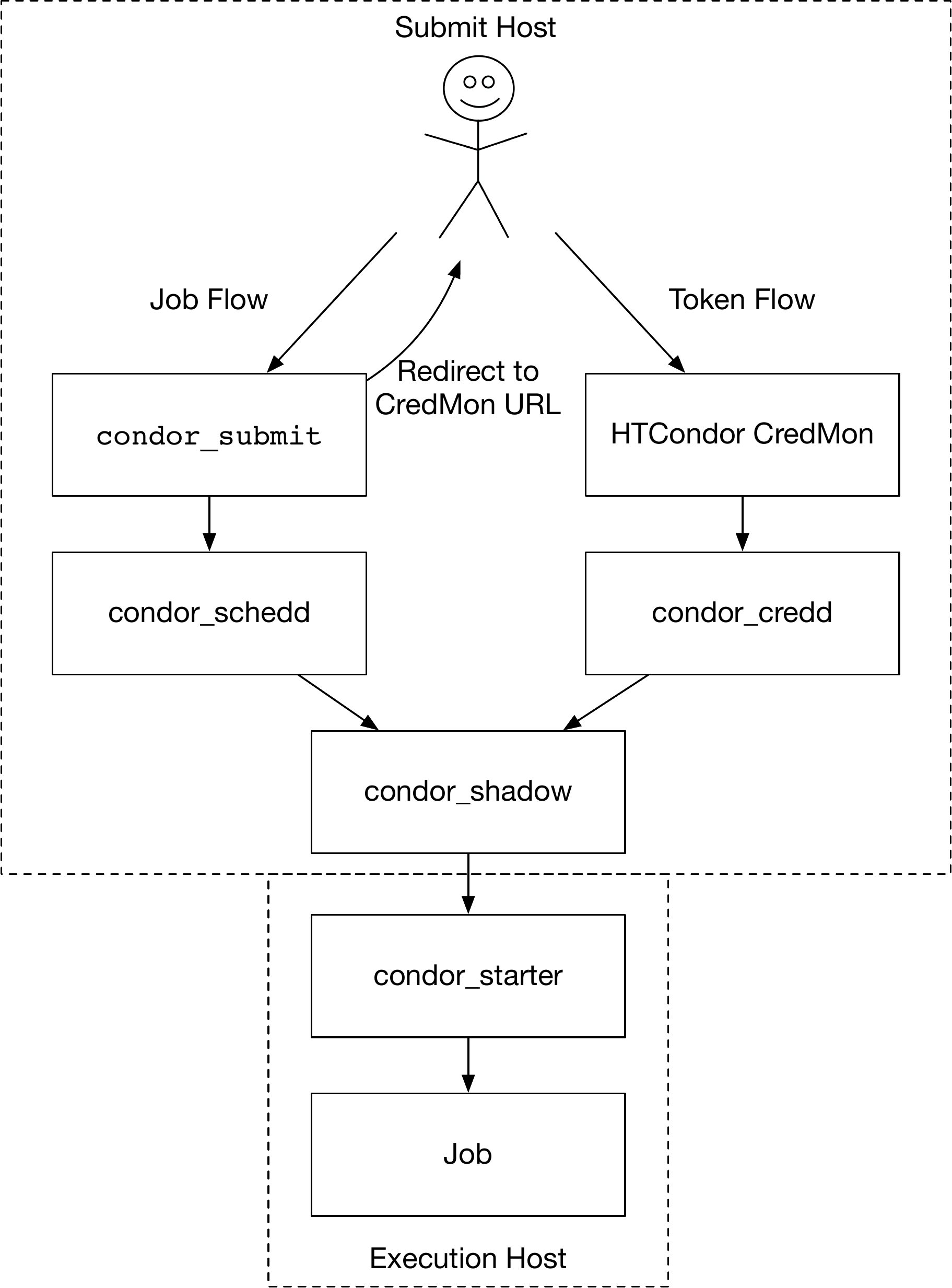}
\caption{Token flow for HTCondor}
\label{fig:htcondortokenflow}
\end{figure}

Figure \ref{fig:htcondortokenflow} displays the job and token flow for a submission to HTCondor.  The user submits their first job with \texttt{condor\_submit}.  The output of \texttt{condor\_submit} will display a URL that the user must visit to acquire credentials.  Once the credentials are acquired, the user resubmits the job with \texttt{condor\_submit}.  The job is transferred to the execution host as well as the credentials.  The credentials are transferred along with the input files before the job begins.

Other services provide similar functionality as Microsoft OneDrive, such as Box \cite{box} and Dropbox \cite{dropbox}.  HTCondor is distributed with a Box \cite{box} client that will use the OAuth token acquired by the HTCondor CredMon.

On the OSG, two data distribution methods are common: StashCache \cite{weitzel2019stashcache} and HTTP Proxies.  StashCache consists of data caches distributed throughout the U.S.  A collection of data origins provide the authoritative source of data to the caches.  Each cache is connected to the U.S. research network with at least 10Gbps connections and several terabytes of storage.

HTTP Proxies where originally designed by the large experiments running on the OSG such as CMS \cite{blumenfeld2008cms}.  A cache is located at every cluster near the worker nodes. The HTTP proxies cache HTTP requests to external resources.

Measurements of StashCache and HTTP Proxies have shown that the performance of either is dependent on the site \cite{weitzel2019stashcache}.  Some sites have significantly better performance with HTTP proxies, while others perform better with StashCache.

A common distribution method for users of cyberinfrastructure is the Globus Online tools \cite{foster2011globus}.  Globus does not provide any storage, rather they provide hosted tools to transfer data between sites.  They do not offer any tools for data distribution to individual jobs though.  Rather, the Globus tools are used for transferring data from storage servers to other storage servers.  There are no cases of using globus online to download data to execution hosts.

\section{Implementation}

Utilizing OneDrive from HTCondor requires the system administrator to:

\begin{enumerate}
    \item Configure OneDrive access in Azure.
    \item Configure the submit host to provide a OAuth client to Azure.
    \item Use the OneDrive OAuth client with HTCondor Integration.
\end{enumerate}

The OneDrive integration uses the Microsoft Graph \cite{msgraph} API in order to access data stored in OneDrive.  Access to the Graph API is configured from within the Azure portal.

\subsection{Configuring Access from Azure}

Configuring the submit host to access OneDrive first starts with registering an OAuth application with Azure Active Directory.  In the Azure portal \cite{azureportal}, go to App Registrations.  Create a new registration.  You will need the hostname of the submit host.  Additionally, the submit host will need an SSL certificate in order for Azure to redirect the credentials to the submit host.  Enter the submit host and the URL that the HTCondor CredMon can receive the credentials.  The URL will be in the form of \texttt{https://<hostname>/return/onedrive}.

Once the application is registered, you must create client credentials.  The client credentials will give you the client secret that the HTCondor CredMon needs in order to commuicate with Azure.  The client secret is used in the OAuth workflow after the user has been redirected back to the submit host.

In order to read and write to OneDrive, the application needs Microsoft Graph permission to the scope \texttt{Files.ReadWrite.All}

\subsection{Configuring the Submit Host}

The submit host needs the configuration from the application registration in Azure.  The following configuration needs to be added:

\begin{itemize}
    \item \textbf{ONEDRIVE\_CLIENT\_ID}: The client ID from the Azure application registration
    \item \textbf{ONEDRIVE\_CLIENT\_SECRET\_FILE}: The path to the client secret from the Azure application registration.  The client secret file should have restricted permissions as it is the single authentication mechanism between the submit host and Microsoft Graph.  With the client secret, another host could impersonate the submit host in communication with Microsoft Graph API.
    \item \textbf{ONEDRIVE\_AUTHORIZATION\_URL}: The authorization URL is always the same\footnote{Authorize URL: https://login.microsoftonline.com/common/oauth2/v2.0/authorize}.
    \item \textbf{ONEDRIVE\_TOKEN\_URL}: The token URL is always the same\footnote{Token URL: https://login.microsoftonline.com/common/oauth2/v2.0/token}.
\end{itemize}

When submitting the job, the user needs to specify that it will need an OneDrive token.  An example submit file is show in Figure \ref{lst:examplesubmit}.

\begin{figure}[ht]
\begin{lstlisting}[frame=single, breaklines=true, postbreak=\mbox{\textcolor{red}{$\hookrightarrow$}\space}]
executable = test.sh
output = out
error = err
log = log
should_transfer_files = YES
when_to_transfer_output = ON_EXIT
transfer_input_files = main
use_oauth_services = onedrive
onedrive_oauth_permissions = Files.ReadWrite.All
queue
\end{lstlisting}
\caption{Example OAuth HTCondor Submit File}
\label{lst:examplesubmit}
\end{figure}

In Figure \ref{lst:examplesubmit}, the lines to note are \texttt{use\_oauth\_services} and \\\texttt{onedrive\_oauth\_permissions}.  The services line tells HTCondor which OAuth services are required for the job.  The services must already be configured on the submit host.  Multiple OAuth services can be configured with a single submit host.

The permissions must be a subset of permissions configured in Azure.  These are scopes will be requested by the HTCondor CredMon.

When a user submits a job requiring a token, they will be prompted to visit the CredMon webservice to authenticate.  The submission prompt is shown in Figure \ref{lst:credentialprompt}.

\begin{figure}[ht]
\begin{lstlisting}[frame=single, breaklines=true, postbreak=\mbox{\textcolor{red}{$\hookrightarrow$}\space}]
$ condor_submit job.condor
Submitting job(s)
Hello, dweitzel.
Please visit: https://<submit_host>/key/86d65e458f6d...
\end{lstlisting}
\caption{Credential Prompt}
\label{lst:credentialprompt}
\end{figure}

\begin{figure}[ht]
    \centering
    \frame{\includegraphics[width=0.5\textwidth]{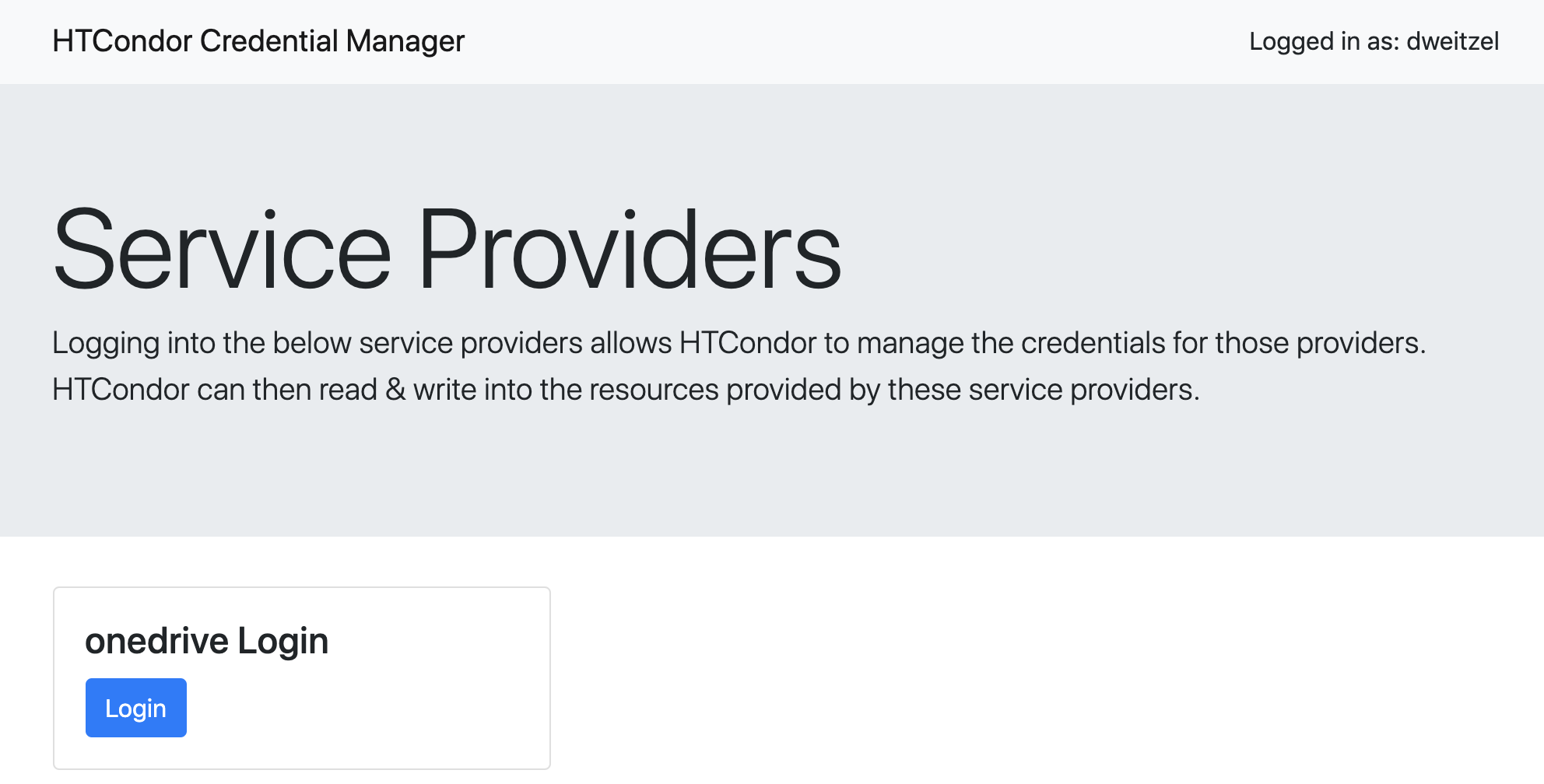}}
    \caption{HTCondor CredMon Web Page}
    \label{fig:credmon-screenshot}
\end{figure}

The user will copy the URL from the prompt and visit the website.  The user clicks on the "login" button that will redirect the user to the OneDrive login screen.  The screen is shown in Figure \ref{fig:credmon-screenshot}.

Once the user has logged into OneDrive, it will redirect the user back to the submit host which will report that the token has been retrieved.  The HTCondor CredMon will receive a special token from OneDrive which, when combined with the client secret from Azure, will be used to retrieve the OAuth refresh token.  After the refresh token is retrieved, it will then be used to receive the access token which is used by the client to download and upload files.

\subsection{Client}
\label{sec:client}

A new client was written to utilize the OAuth tokens received from Azure \cite{derek_weitzel_2019_3265184}.  Other clients for OneDrive also use OAuth tokens, but request the token on their own.  The client that works with HTCondor must accept an existing token and not require re-authorization when running on remote hosts.  Additionally, the clients cannot parse the token format that HTCondor creates.

The new client is written in the Go \cite{golang} programming language.  Go executables are statically compiled with their dependencies and therefore require no extra software or libraries on remote resources.

The client is called on the command line as shown in Figure \ref{lst:clientcommandline}.

\begin{figure}[ht]
\begin{lstlisting}[frame=single, breaklines=true, postbreak=\mbox{\textcolor{red}{$\hookrightarrow$}\space}]
./onedrive onedrive:///file.txt ./
\end{lstlisting}
\caption{Client command line}
\label{lst:clientcommandline}
\end{figure}

To download a file, the client first retrieves the access token.  The token stored on the remote resource in a directory pointed to by the \texttt{\_CONDOR\_CREDS} environment variable.  The token file will be named \texttt{onedrive.use}.  The token file contains a JSON data structure with information about the token, as well as the token itself.  The client only needs the access token.

The client then creates the Azure Graph URL request to download the file contents.  The token gathered from the JSON will be inserted into the \texttt{Authorization} line of the HTTP headers.  An example request is shown in Figure \ref{lst:graphheaders}.

\begin{figure}[ht]
\begin{lstlisting}[frame=single, breaklines=true, postbreak=\mbox{\textcolor{red}{$\hookrightarrow$}\space}]
GET /v1.0/me/drive/root:/file.txt:/content HTTP/1.0
Host: graph.microsoft.com
Authorization: Bearer <encoded_token>
\end{lstlisting}
\caption{HTTP Headers for Azure Graph OneDrive request}
\label{lst:graphheaders}
\end{figure}

The file is downloaded to the destination file or directory that was specified on the command line.  The download is secured with HTTPS connections.  The token and the download are both secure from third party interception.

\section{Evaluation}

I designed an experiment based on real usage reported on the Open Science Grid's data distribution framework, StashCache.  The usage was reported in a paper about StashCache \cite{weitzel2019stashcache}.

Test files were created by measuring the usage of OSG's StashCache.  The test file sizes cover the percentails of usage.  Table \ref{tab:stashcachepercentiles} shows the test file sizes.  I uploaded files of the sizes shown in Table \ref{tab:stashcachepercentiles} to OneDrive.

\rowcolors{2}{gray!25}{white}
\begin{table}[ht]
    \centering
    \caption{StashCache File Size Percentiles}
    \begin{tabular}{l|l}
        \rowcolor{gray!50}
        \textbf{Percentile} & \textbf{Filesize} \\ \hline
        1 & 5.797KB \\
        5 & 22.801MB \\
        25 & 170.131MB \\ 
        50 & 467.852MB \\
        75 & 493.337MB \\
        95 & 2.335GB \\
        99 & 2.335GB
    \end{tabular}
    \label{tab:stashcachepercentiles}
\end{table}

I created tests that will download each of the files and measure the download speed.  For HTTP and StashCache, we perform two downloads in order to guarantee that the second download is from the cache.  Similarly for OneDrive, we download the file twice in case the downloads are cached inside of OneDrive.

The test used the OneDrive credential integration as well as the client discussed in Section \ref{sec:client}.  I created a HTCondor submit file which would copy the client to the worker node in order to download the files.  A HTCondor Directed Acyclic Graph (DAG) \cite{frey2002condor} workflow was developed to submit to each of the clusters, one at a time.  We downloaded from each site one at a time to be consistent with the StashCache paper and to not encounter bottlenecks in the infrastructure.

I chose to run against the same sites as the StashCache paper, the top 4 sites providing opportunistic computing in the last six months on the OSG.  I did not run at University of Nebraska--Lincoln as Nebraska was not capable of running opportunistic jobs at the time of writing.  The sites we did run at are Syracuse University, University of Colorado, Bellarmine University, and the University of Chicago.

I will compare the download performance of OneDrive with the performance of StashCache and HTTP proxies.  These transfer methods are the most commonly used data distribution methods on the OSG.  The measurements of the StashCache and HTTP proxies are from the StashCache \cite{weitzel2019stashcache} paper.

\section{Results and Discussion}

The results of the download tests are shown in several following figures.  In the figures, each of the file sizes are compared together.  The transfer speed in megabits per second (mbps) is compared.


\begin{figure}[ht]
\centering
        \includegraphics[width=0.45\textwidth]{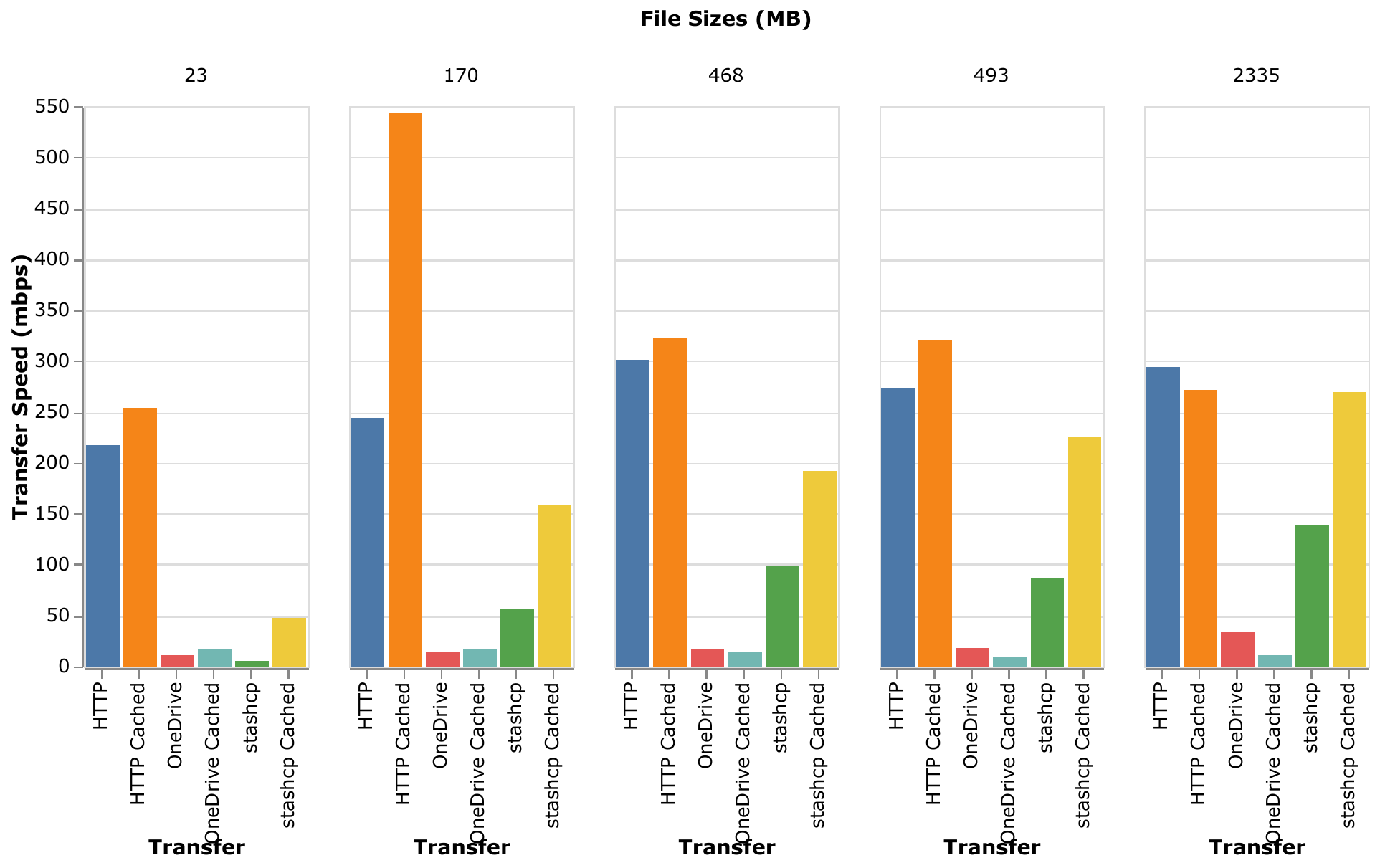}
        \caption{Syracuse Download Comparison}
        \label{fig:syracuse-download}
    \end{figure}
    \begin{figure}[ht]
        \centering
        \includegraphics[width=0.45\textwidth]{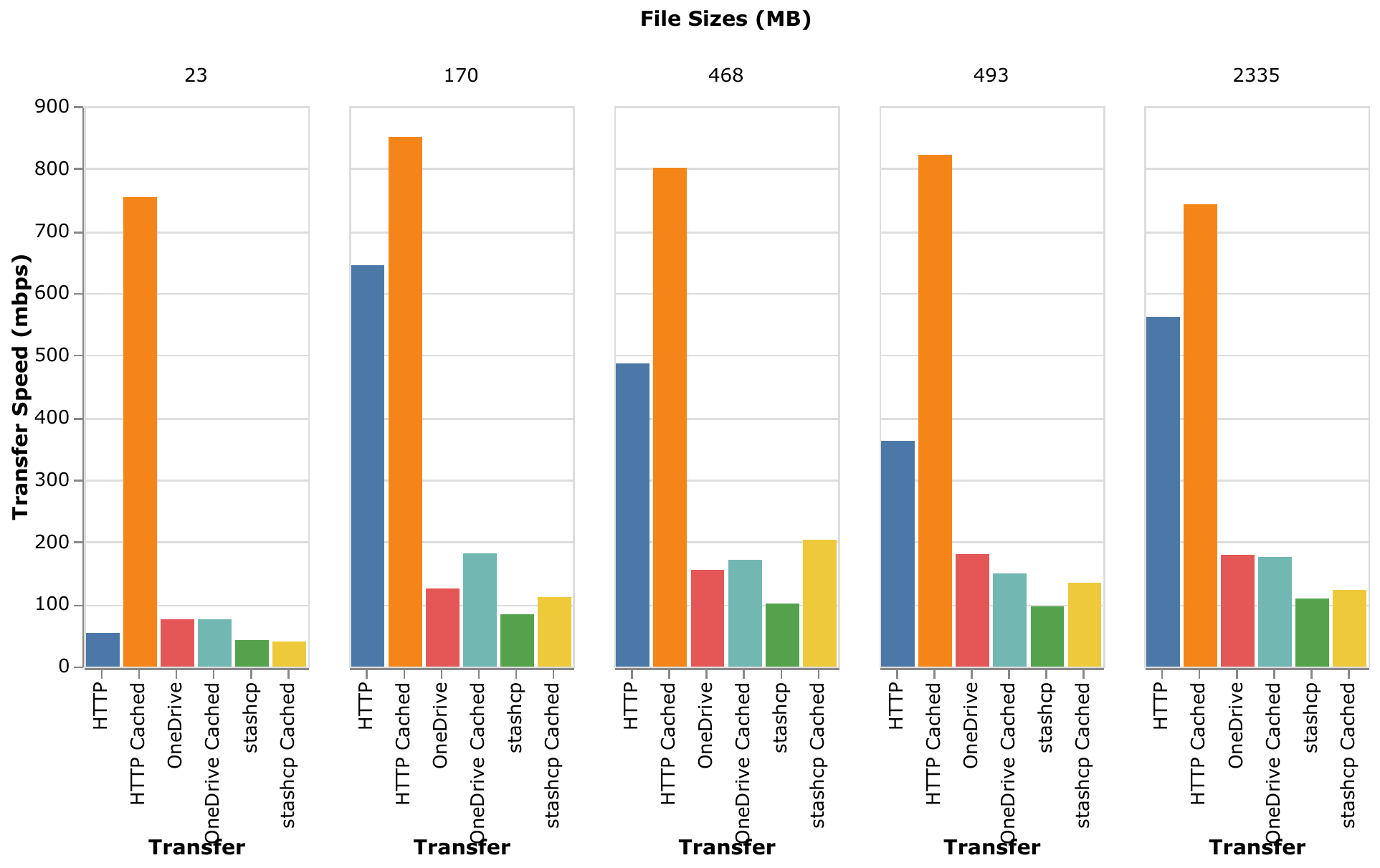}
        \caption{Colorado Download Comparison}
        \label{fig:colorado-download}
    \end{figure}

As you can see in Figure \ref{fig:syracuse-download}, OneDrive transfers are always slower than the HTTP or StashCache for Syracuse cluster.  OneDrive is not optimized for transfers, and does not have regional caches similar to StashCache.  HTTP transfers may have a cache very near the worker nodes.  Additionally, those clusters are connected to research networks, such as Internet 2 \cite{internet2}, that are optimized for site to site transfers between research institutions.  The caches and the origins of data for the HTTP and StashCache data are also connected to the research networks.

In contrast, Figure \ref{fig:colorado-download} shows that for Colorado StashCache performs worse than OneDrive.  The site configuration at Colorado may limit the bandwidth to the research network.  It is clear from the download graph that the HTTP downloads from Colorado are prioritized somehow, while StashCache and OneDrive downloads are not.

\begin{figure}[ht]
    \centering
    \includegraphics[width=0.45\textwidth]{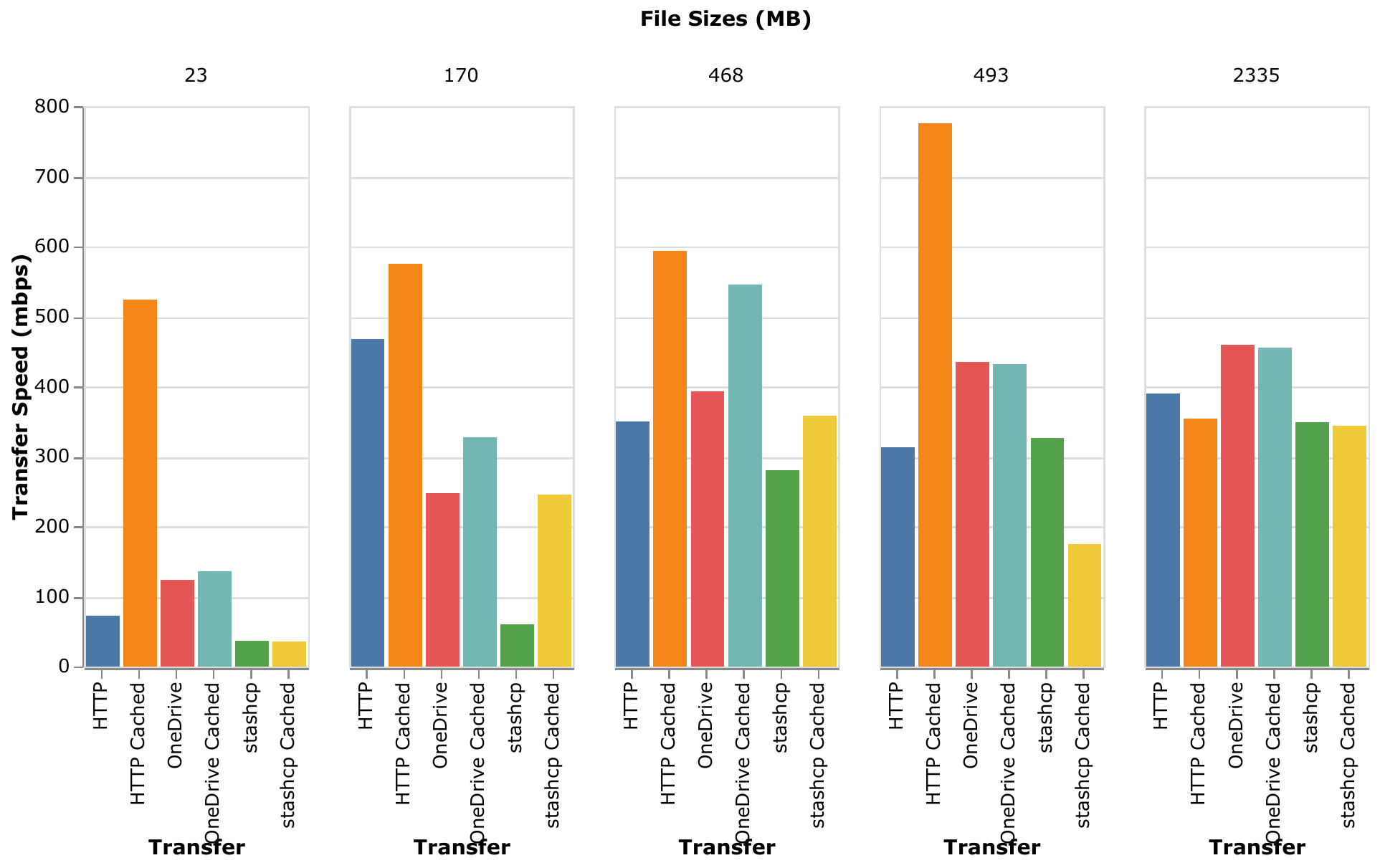}
    \caption{Bellarmine Download Speed Comparison}
    \label{fig:bell-download}
\end{figure}

\rowcolors{2}{gray!25}{white}
\begin{table*}[!htb]
    \centering
    \caption{Comparing OneDrive with StashCache or HTTP.  This is the percent difference between the OneDrive and StashCache or HTTP.  Positive values mean that OneDrive was faster at downloading the test file.}
    \begin{tabular}{lrrrr}
        \rowcolor{gray!50}
        \textbf{Site} & \textbf{HTTP 22.8MB} & \textbf{HTTP 2.3 GB} & \textbf{StashCache 22.8MB} & \textbf{StashCache 2.3GB} \\ \hline
        Syracuse  & -80.0\%& -77.8\% & 7.3\% & -77.6\% \\
        Colorado  & -89.9\% & -75.9\% & 89.4\% &  46.1\% \\
        Bellarmine & -76.4\% &  29.7\% & 248.1\% & 33.8\% \\ 
        Chicago & -85.5\% &  4.1\% & -41.6\% & 36.0\% \\ \hline
        Average & -82.9\% & -29.9\% & 75.8\% &  9.5\% \\ \hline
    \end{tabular}

    \label{tab:comparisons}
\end{table*}

As noted in the StashCache paper \cite{weitzel2019stashcache}, sites differ dramatically in their data deliver performance.  Figure \ref{fig:bell-download} shows the download performance of the Bellarmine cluster.  You can see that OneDrive performs much better than it does at both Syracuse and Colorado.  In fact, OneDrive performed faster downloads for the 2.3GB file than either HTTP or StashCache.

\begin{figure}[ht]
    \centering
    \includegraphics[width=0.45\textwidth]{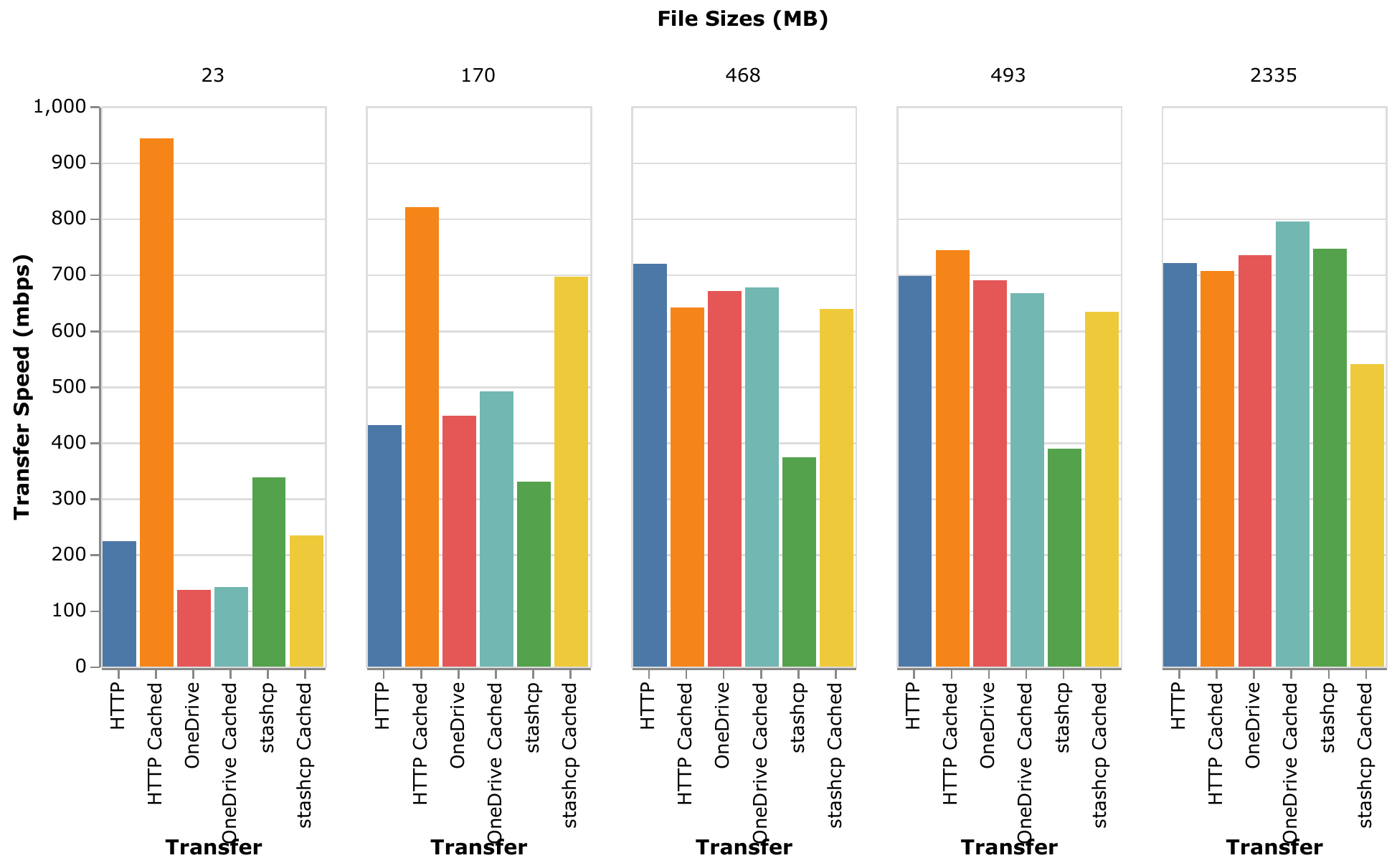}
    \caption{University of Chicago Download Speed Comparison}
    \label{fig:mwt2-download}
\end{figure}

Figure \ref{fig:mwt2-download} shows the download speed comparison for the University of Chicago.  OneDrive download speed is comparable to HTTP and StashCache.

Table \ref{tab:comparisons} Compares the performance of the different data distribution methods for the the 22MB and 2.3GB test files.  On average, OneDrive performs better than StashCache, but worse than HTTP.

We downloaded the file from OneDrive twice in order to detect any affect of caching within the OneDrive architecture.  But, no effect of caching was detected when running the tests.

No issues where observed while performing the experiments.  OneDrive downloads all completed successfully with the client.


\section{Conclusions \& Future Work}

I formulated that OneDrive would perform worse on the OSG than other data distribution methods such as HTTP and StashCache.  But, my experiments proved that OneDrive can perform comparable in performance to HTTP and StashCache, and at some sites, even perform better than other data delivery methods.

Some sites perform very well in OneDrive downloads, while others perform poorly.  At Syracuse for example, OneDrive downloads performed worse than HTTP or StashCache.  It is not clear why this is the case.  It is possible that Syrcause does not have a fast connection to the Microsoft network, especially compared to other sites.  More study of this inconsistent download performance is warranted, including running these tests across more sites.

During the experiments, we were unable to determine where the OneDrive downloads are coming from.  This could have been useful in comparing the download performance of OneDrive.  It's possible that distance from the datacenter hosting the OneDrive data could have decreased the download speed.

In the future, we will add writeback capabilities to the client.  Writeback to OneDrive storage will allow users to automatically send their results to any clients that are syncing the OneDrive folder.
	
	\bibliographystyle{ACM-Reference-Format}
	\bibliography{weitzel_onedrive}


\begin{thebibliography}{18}


\ifx \showCODEN    \undefined \def \showCODEN     #1{\unskip}     \fi
\ifx \showDOI      \undefined \def \showDOI       #1{#1}\fi
\ifx \showISBNx    \undefined \def \showISBNx     #1{\unskip}     \fi
\ifx \showISBNxiii \undefined \def \showISBNxiii  #1{\unskip}     \fi
\ifx \showISSN     \undefined \def \showISSN      #1{\unskip}     \fi
\ifx \showLCCN     \undefined \def \showLCCN      #1{\unskip}     \fi
\ifx \shownote     \undefined \def \shownote      #1{#1}          \fi
\ifx \showarticletitle \undefined \def \showarticletitle #1{#1}   \fi
\ifx \showURL      \undefined \def \showURL       {\relax}        \fi
\providecommand\bibfield[2]{#2}
\providecommand\bibinfo[2]{#2}
\providecommand\natexlab[1]{#1}
\providecommand\showeprint[2][]{arXiv:#2}

\bibitem[\protect\citeauthoryear{Aad, Butterworth, Thion, Bratzler, Ratoff,
  Nickerson, Seixas, Grabowska-Bold, Meisel, Lokwitz, et~al\mbox{.}}{Aad
  et~al\mbox{.}}{2008}]%
        {aad2008atlas}
\bibfield{author}{\bibinfo{person}{Georges Aad}, \bibinfo{person}{JM
  Butterworth}, \bibinfo{person}{J Thion}, \bibinfo{person}{U Bratzler},
  \bibinfo{person}{PN Ratoff}, \bibinfo{person}{RB Nickerson},
  \bibinfo{person}{JM Seixas}, \bibinfo{person}{I Grabowska-Bold},
  \bibinfo{person}{F Meisel}, \bibinfo{person}{S Lokwitz}, {et~al\mbox{.}}}
  \bibinfo{year}{2008}\natexlab{}.
\newblock \showarticletitle{The ATLAS experiment at the CERN large hadron
  collider}.
\newblock \bibinfo{journal}{\emph{Jinst}}  \bibinfo{volume}{3}
  (\bibinfo{year}{2008}), \bibinfo{pages}{S08003}.
\newblock


\bibitem[\protect\citeauthoryear{Blumenfeld, Dykstra, Lueking, and
  Wicklund}{Blumenfeld et~al\mbox{.}}{2008}]%
        {blumenfeld2008cms}
\bibfield{author}{\bibinfo{person}{Barry Blumenfeld}, \bibinfo{person}{David
  Dykstra}, \bibinfo{person}{Lee Lueking}, {and} \bibinfo{person}{Eric
  Wicklund}.} \bibinfo{year}{2008}\natexlab{}.
\newblock \showarticletitle{CMS conditions data access using FroNTier}. In
  \bibinfo{booktitle}{\emph{Journal of Physics: Conference Series}},
  Vol.~\bibinfo{volume}{119}. IOP Publishing, \bibinfo{pages}{072007}.
\newblock


\bibitem[\protect\citeauthoryear{Box}{Box}{2019}]%
        {box}
\bibfield{author}{\bibinfo{person}{Box}.} \bibinfo{year}{2019}\natexlab{}.
\newblock \bibinfo{booktitle}{\emph{Box}}.
\newblock
\urldef\tempurl%
\url{https://www.box.com/}
\showURL{%
\tempurl}


\bibitem[\protect\citeauthoryear{Chatrchyan, de~Wolf, et~al\mbox{.}}{Chatrchyan
  et~al\mbox{.}}{2008}]%
        {chatrchyan2008cms}
\bibfield{author}{\bibinfo{person}{Serguei Chatrchyan}, \bibinfo{person}{EA de
  Wolf}, {et~al\mbox{.}}} \bibinfo{year}{2008}\natexlab{}.
\newblock \showarticletitle{The CMS experiment at the CERN LHC}.
\newblock \bibinfo{journal}{\emph{Journal of instrumentation.-Bristol, 2006,
  currens}}  \bibinfo{volume}{3} (\bibinfo{year}{2008}),
  \bibinfo{pages}{S08004--1}.
\newblock


\bibitem[\protect\citeauthoryear{Dropbox}{Dropbox}{2019}]%
        {dropbox}
\bibfield{author}{\bibinfo{person}{Dropbox}.} \bibinfo{year}{2019}\natexlab{}.
\newblock \bibinfo{booktitle}{\emph{Dropbox}}.
\newblock
\urldef\tempurl%
\url{https://www.dropbox.com/}
\showURL{%
\tempurl}


\bibitem[\protect\citeauthoryear{Foster}{Foster}{2011}]%
        {foster2011globus}
\bibfield{author}{\bibinfo{person}{Ian Foster}.}
  \bibinfo{year}{2011}\natexlab{}.
\newblock \showarticletitle{Globus Online: Accelerating and democratizing
  science through cloud-based services}.
\newblock \bibinfo{journal}{\emph{IEEE Internet Computing}}
  \bibinfo{volume}{15}, \bibinfo{number}{3} (\bibinfo{year}{2011}),
  \bibinfo{pages}{70--73}.
\newblock


\bibitem[\protect\citeauthoryear{Frey}{Frey}{2002}]%
        {frey2002condor}
\bibfield{author}{\bibinfo{person}{James Frey}.}
  \bibinfo{year}{2002}\natexlab{}.
\newblock \bibinfo{title}{Condor DAGMan: Handling inter-job dependencies}.
\newblock
\newblock


\bibitem[\protect\citeauthoryear{GO}{GO}{2019}]%
        {golang}
\bibfield{author}{\bibinfo{person}{GO}.} \bibinfo{year}{2019}\natexlab{}.
\newblock \bibinfo{booktitle}{\emph{The Go Programming Language}}.
\newblock
\urldef\tempurl%
\url{https://golang.org/}
\showURL{%
\tempurl}


\bibitem[\protect\citeauthoryear{Internet2}{Internet2}{2019}]%
        {internet2}
\bibfield{author}{\bibinfo{person}{Internet2}.}
  \bibinfo{year}{2019}\natexlab{}.
\newblock \bibinfo{booktitle}{\emph{Internet2}}.
\newblock
\urldef\tempurl%
\url{https://www.internet2.edu/}
\showURL{%
\tempurl}


\bibitem[\protect\citeauthoryear{Microsoft}{Microsoft}{2019a}]%
        {azureportal}
\bibfield{author}{\bibinfo{person}{Microsoft}.}
  \bibinfo{year}{2019}\natexlab{a}.
\newblock \bibinfo{booktitle}{\emph{Microsoft Azure Portal}}.
\newblock
\urldef\tempurl%
\url{https://portal.azure.com}
\showURL{%
\tempurl}


\bibitem[\protect\citeauthoryear{Microsoft}{Microsoft}{2019b}]%
        {msgraph}
\bibfield{author}{\bibinfo{person}{Microsoft}.}
  \bibinfo{year}{2019}\natexlab{b}.
\newblock \bibinfo{booktitle}{\emph{Microsoft Graph}}.
\newblock
\urldef\tempurl%
\url{https://developer.microsoft.com/en-us/graph}
\showURL{%
\tempurl}


\bibitem[\protect\citeauthoryear{Microsoft}{Microsoft}{2019c}]%
        {onedrive}
\bibfield{author}{\bibinfo{person}{Microsoft}.}
  \bibinfo{year}{2019}\natexlab{c}.
\newblock \bibinfo{booktitle}{\emph{Microsoft OneDrive}}.
\newblock
\urldef\tempurl%
\url{https://onedrive.live.com/about/en-us/}
\showURL{%
\tempurl}


\bibitem[\protect\citeauthoryear{Pordes, Petravick, Kramer, Olson, Livny, Roy,
  Avery, Blackburn, Wenaus, W{\"u}rthwein, et~al\mbox{.}}{Pordes
  et~al\mbox{.}}{2007}]%
        {pordes2007open}
\bibfield{author}{\bibinfo{person}{Ruth Pordes}, \bibinfo{person}{Don
  Petravick}, \bibinfo{person}{Bill Kramer}, \bibinfo{person}{Doug Olson},
  \bibinfo{person}{Miron Livny}, \bibinfo{person}{Alain Roy},
  \bibinfo{person}{Paul Avery}, \bibinfo{person}{Kent Blackburn},
  \bibinfo{person}{Torre Wenaus}, \bibinfo{person}{Frank W{\"u}rthwein},
  {et~al\mbox{.}}} \bibinfo{year}{2007}\natexlab{}.
\newblock \showarticletitle{The open science grid}. In
  \bibinfo{booktitle}{\emph{Journal of Physics: Conference Series}},
  Vol.~\bibinfo{volume}{78}. IOP Publishing, \bibinfo{pages}{012057}.
\newblock


\bibitem[\protect\citeauthoryear{Thain, Tannenbaum, and Livny}{Thain
  et~al\mbox{.}}{2005}]%
        {thain2005distributed}
\bibfield{author}{\bibinfo{person}{Douglas Thain}, \bibinfo{person}{Todd
  Tannenbaum}, {and} \bibinfo{person}{Miron Livny}.}
  \bibinfo{year}{2005}\natexlab{}.
\newblock \showarticletitle{Distributed computing in practice: the Condor
  experience}.
\newblock \bibinfo{journal}{\emph{Concurrency and computation: practice and
  experience}} \bibinfo{volume}{17}, \bibinfo{number}{2-4}
  (\bibinfo{year}{2005}), \bibinfo{pages}{323--356}.
\newblock


\bibitem[\protect\citeauthoryear{Tridgell, Mackerras, et~al\mbox{.}}{Tridgell
  et~al\mbox{.}}{1996}]%
        {tridgell1996rsync}
\bibfield{author}{\bibinfo{person}{Andrew Tridgell}, \bibinfo{person}{Paul
  Mackerras}, {et~al\mbox{.}}} \bibinfo{year}{1996}\natexlab{}.
\newblock \showarticletitle{The rsync algorithm}.
\newblock  (\bibinfo{year}{1996}).
\newblock


\bibitem[\protect\citeauthoryear{Weitzel}{Weitzel}{2019}]%
        {derek_weitzel_2019_3265184}
\bibfield{author}{\bibinfo{person}{Derek Weitzel}.}
  \bibinfo{year}{2019}\natexlab{}.
\newblock \bibinfo{title}{{djw8605/onedrive-oauth: First release of OneDrive
  oauth}}.
\newblock
\newblock
\urldef\tempurl%
\url{https://doi.org/10.5281/zenodo.3265184}
\showDOI{\tempurl}


\bibitem[\protect\citeauthoryear{Weitzel, Zvada, Vukotic, Gardner, Bockelman,
  Rynge, Hernandez, Lin, and Selmeci}{Weitzel et~al\mbox{.}}{2019}]%
        {weitzel2019stashcache}
\bibfield{author}{\bibinfo{person}{Derek Weitzel}, \bibinfo{person}{Marian
  Zvada}, \bibinfo{person}{Ilija Vukotic}, \bibinfo{person}{Rob Gardner},
  \bibinfo{person}{Brian Bockelman}, \bibinfo{person}{Mats Rynge},
  \bibinfo{person}{Edgar~Fajardo Hernandez}, \bibinfo{person}{Brian Lin}, {and}
  \bibinfo{person}{Matyas Selmeci}.} \bibinfo{year}{2019}\natexlab{}.
\newblock \showarticletitle{StashCache: A Distributed Caching Federation for
  the Open Science Grid}.
\newblock \bibinfo{journal}{\emph{Proceedings of the Practice and Experience on
  Advanced Research Computing}} (\bibinfo{year}{2019}).
\newblock


\bibitem[\protect\citeauthoryear{Withers, Bockelman, Weitzel, Brown, Gaynor,
  Basney, Tannenbaum, and Miller}{Withers et~al\mbox{.}}{2018}]%
        {withers2018scitokens}
\bibfield{author}{\bibinfo{person}{Alex Withers}, \bibinfo{person}{Brian
  Bockelman}, \bibinfo{person}{Derek Weitzel}, \bibinfo{person}{Duncan Brown},
  \bibinfo{person}{Jeff Gaynor}, \bibinfo{person}{Jim Basney},
  \bibinfo{person}{Todd Tannenbaum}, {and} \bibinfo{person}{Zach Miller}.}
  \bibinfo{year}{2018}\natexlab{}.
\newblock \showarticletitle{SciTokens: Capability-Based Secure Access to Remote
  Scientific Data}. In \bibinfo{booktitle}{\emph{Proceedings of the Practice
  and Experience on Advanced Research Computing}}. ACM, \bibinfo{pages}{24}.
\newblock


\end{thebibliography}

\end{document}